# A Method of Rapid Quantification of Patient-Specific Organ Dose for CT Using Coupled Deep-Learning based Multi-Organ Segmentation and GPU-accelerated Monte Carlo Dose Computing

Zhao Peng[1,2,3], Xi Fang[1], Pingkun Yan[1], Hongming Shan[1], Tianyu Liu[2], Xi Pei[3,4], Ge Wang[1], Bob Liu[5], Mannudeep K. Kalra[5], X. George Xu[1,2,3,*]

[1] Department of Biomedical Engineering, Rensselaer Polytechnic Institute, Troy, NY 12180, USA

[2] Department of Mechanical, Aerospace and Nuclear Engineering, Rensselaer Polytechnic Institute, Troy, NY 12180, USA

[3] Department of Engineering and Applied Physics, University of Science and Technology of China, Hefei, Anhui 230026, China

[4] Anhui Wisdom Technology Company Limited, Hefei, Anhui, 238000, China

[5] Department of Radiology, Massachusetts General Hospital, Boston, MA, USA

*Corresponding Author:

Prof. X. George Xu, Ph.D., FAAPM, FANS, and FHPS

Edward E. Hood Endowed Chair Professor of Engineering

Rensselaer Polytechnic Institute

Troy, New York 12180, USA

Tel: 518-276-4014

Email: xug2@rpi.edu

**ABSTRACT**

**Purpose:** The ability to obtain patient-specific organ doses for CT will open the door to new applications such as personalized selection of scan factors and individualized risk assessment, leading to the ultimate goal of achieving low-dose and optimized CT imaging. One technical barrier to advancing CT dosimetry has been the lack of computational tools for automatic patient-specific multi-organ segmentation of CT images, coupled with rapid organ dose quantification. This study aims to demonstrate the feasibility of combining deep-learning algorithms for automatic segmentation of radiosensitive organs from CT images and GPU-based Monte Carlo rapid organ dose calculation.

**Methods:** A deep convolutional neural network (CNN) based on the U-Net for organ segmentation is developed and trained to automatically delineate radiosensitive organs from CT images. Two databases are used: the Lung CT Segmentation Challenge 2017 (LCTSC) dataset that contains 60 thoracic CT scan patients each with 5 segmented organs, and the Pancreas-CT (PCT) dataset that contains 43 abdominal CT scan patients each with 8 segmented organs. A five-fold cross-validation of the new method is performed on both sets of data. Dice Similarity Coefficient (DSC) is used to evaluate the segmentation performance against the ground truth. A GPU-based Monte Carlo dose code, ARCHER, is used to calculate patient-specific CT organ doses. The proposed method is tested in terms of Relative Dose Error (RDE). To demonstrate the potential improvement of the new methods, organ dose results are compared against those obtained for population-average phantoms used in an off-line dose reporting software, VirtualDose, at Massachusetts General Hospital.

**Results:** For the group of 60 patients from LCTSC dataset, the median DSCs are found to be 0.97 (right lung), 0.96 (left lung), 0.93 (heart), 0.88 (spinal cord) and 0.78 (esophagus). For the group of 43 patients from PCT dataset, the median DSCs are found to be 0.96 (spleen), 0.96 (liver), 0.95 (left kidney), 0.89 (stomach), 0.87 (gall bladder), 0.79 (pancreas), 0.74 (esophagus), and 0.64 (duodenum). Comparing with the organ dose results from population-averaged phantoms, the new patient-specific method achieved the smaller RDE range on all organs: -4.3%~1.5% (vs -31.5%~33.9%) for the lung, -7.0%~2.3% (vs -15.2%~125.1%) for the heart, -18.8%~40.2% (vs -10.3%~124.1%) for the esophagus, -5.6%~1.6% (vs -20.3%~57.4%) for the spleen, -4.5%~4.6% (vs -19.5%~61.0%) for the pancreas, -2.3%~4.4% (vs -37.8%~75.8%) for the left kidney, -14.9%~5.4% (vs -39.9% ~14.6%) for the gallbladder, -0.9%~1.6% (vs -30.1%~72.5%) for the liver, and -23.0%~11.1% (vs -52.5%~-1.3%) for the stomach. The trained automatic segmentation tool takes less than 5 seconds in a patient for all 103 patients in the dataset. The Monte Carlo radiation dose calculations performed in parallel with the segmentation using the GPU-accelerated ARCHER code takes less than 4 seconds in a patient to achieve <0.5% statistical uncertainty in all organ doses for all 103 patients in the database.

**Conclusion:** It is feasible to perform streamlined automatic organ segmentation from patient-specific CT images and rapid GPU-based Monte Carlo dose quantification with clinically acceptable accuracy and efficiency.

# 1. INTRODUCTION

In the United States, the number of CT examinations increased 20 folds between the 1980s and 2010s, owing to rapid improvements in multi-detector CT (MDCT) technologies [1-3]. In 2018, about 88.7 million CT examinations were performed in the United States alone, which represented a substantial increase from 21 million exams in 1995 [4]. The abdomen and chest regions represent the most frequently scanned body regions accounting for more than a third of all CT examinations. Given the rising use of CT and concerns over associated radiation risks, the American College of Radiology (ACR) has called for more research and development in patient-specific dose quantification, scanner optimization, and protocol comparison [1].

The currently displayed CTDIvol and DLP are technical dose descriptors and do not represent or take into account patient body habitus (size or shape), attenuation, scanned anatomy, age, gender, or actually absorbed radiation doses [5]. Although CTDIvol and DLP provide a good method to compare scanners and scan protocols, they cannot be used to compare, monitor, or assess patient-specific radiation doses from CT. For this reason, size-specific dose estimates (SSDE) have been recommended as an improved approach that takes into account for patient body habitus [6]. Many methods of generating organ-specific dose databases have been reported [7-15]. These methods require the Monte Carlo simulations of CT scanner components as well as radiation interactions with whole-body computational phantoms that contain organs/tissues explicitly defined in tiny voxels in accordance with the "Reference-Man" concept – population averaged anatomical parameters originally defined for radiation protection purposes [16]. However, the process that was required to create such whole-body phantoms is prohibitively complex for routine analysis of patient-specific images. As a result, most clinical end-users can only perform CT organ dose assessment using "off-line" software tools, such as VirtualDose [15], that are based on databases pre-calculated from a library of population-averaged phantoms. In contrast, in radiation cancer treatment, delineation of the target volume and adjacent healthy organs at risk (OARs) is performed routinely using patient-specific images, followed by rapid dose calculation and inverse treatment plan optimization to minimize normal tissue complication probability (NTCP) [17-19]. Recently, GPU-based Monte Carlo dose computing codes, including ARCHER [20, 21], can achieve clinically acceptable speeds for both patient CT imaging dose assessment and for treatment planning. Therefore, patient-specific organ dose computing methods already exist.

A prospective patient-specific organ dose method will be a game changer in CT and can help extend the existing tube-current modulation techniques by taking full advantage of organ localization and distribution of organ doses. It can also aid imaging physicians make informed and prospective decisions regarding delivery of doses based on the clinical question, expected disease distribution and organ dose distribution. Such prospective decisions regarding radiation dose delivery from CT can help usher personalized scan protocols with truly organ dose-modulated techniques.

Obviously, one technical barrier is the lack of organ localization/segmentation as well as rapid organ dose prediction tools.

Here we are interested in organ dose quantification for a set of radiosensitive organs in every patient undergoing CT scans, for radiation protection purposes (instead of cancer treatment planning purposes). Segmentation of radiosensitive organ volumes from CT images has long been a challenging task to the medical physics community [22]. Manual organ segmentation is labor-intensive and user-dependent, making the approach impractical for clinical applications involving patient-specific images. Until recently, methods of automatic segmentation of organs relied on low-level image features that require strong prior knowledge about the anatomical structures and are insufficient [23]. The advent of deep learning methods involving convolutional neural network (CNN) has brought an unprecedented level of innovation to the field of image segmentation [24-27]. The state-of-the-art models in organ segmentation are variants of encoder-decoder architecture such as the Fully Convolutional Networks (FCN) [28] and U-Net [29]. However, these models are trained usually for specific organs and cannot be easily extended to multi-organ segmentation needed for organ dosimetry for CT. Recently, Trullo et.al. [30] used a modified 2D FCN to segment four organs at risks from CT images and apply Conditional Random Fields (CRF) to further improve the segmentation performance. Gibson et al. [31] applied a 3D Dense V-Network to segment eight organs from CT images for navigation in endoscopic pancreatic and biliary procedures. However, these studies did not perform organ dose calculations for patients who receive the CT scans. Recent studies by other groups that did consider CT organs dose evaluations employed traditional organ segmentation algorithm such as feature-based or atlas-based methods [32, 33]. Finally, without the necessary accuracy and efficiency, patient-specific dosimetry tools would not become a viable part of the clinical workflow.

This study [34, 35] aims to demonstrate the feasibility of a streamlined fast patient-specific CT organ dose assessment method that performs segmentation of multiple organs from patient-specific CT images using deep CNN algorithms and GPU-accelerated Monte Carlo dose calculations using the ARCHER code in a parallel computational workflow as illustrated in Fig. 1. This is the first study to combine these two tools to achieve the computational accuracy and efficiency required for routine clinical applications. In subsequent sections, we describe steps and methods, summarize results, discuss limitations before drawing conclusions.

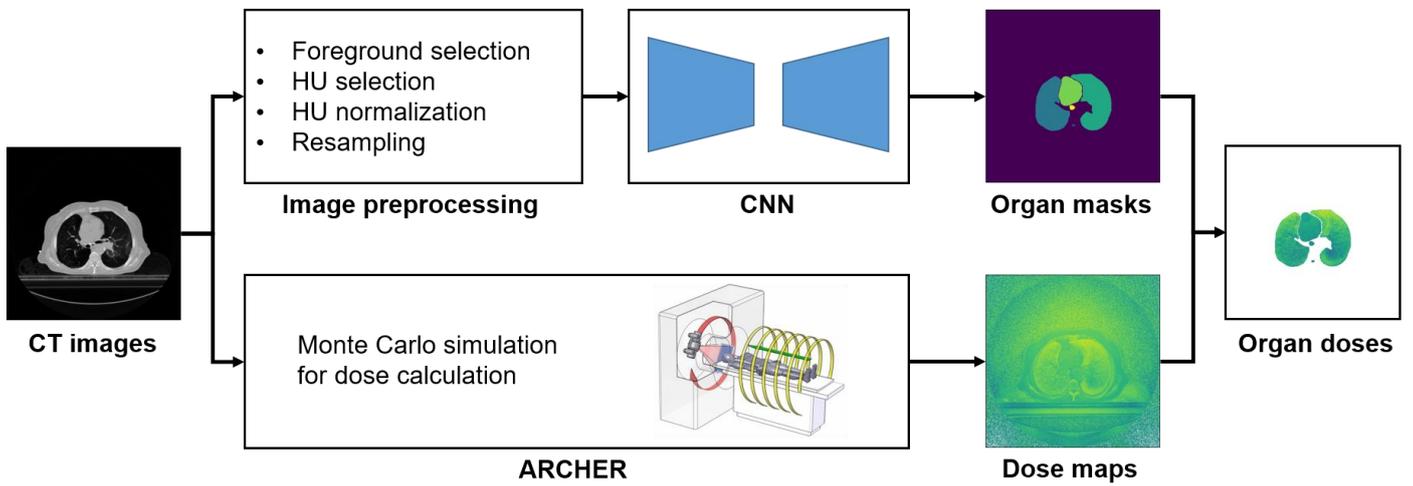

**Fig. 1.** The overall parallel computational process of the method of patient-specific organ dose assessment for CT using multi-organ segmentation CNN algorithms coupled with a GPU Monte Carlo dose engine, ARCHER

## 2. MATERIALS AND METHODS

### 2.A Organ segmentation

### 2.A.1 Datasets and image preprocessing

In this study, two publically available datasets were used: (1) The 2017 Lung CT Segmentation Challenge (LCTSC) [36-38], which contains 60 thoracic CT scan patients with 5 segmented organs (left lung, right lung, heart, spinal cord, and esophagus). (2) Pancreas-CT (PCT), which contains 43 abdominal contrast enhanced CT scan patients with 8 segmented organs (the spleen, left kidney, gallbladder, esophagus, liver, stomach, pancreas and duodenum) [26, 31, 38, 39]. For each patient in these two datasets, the Hounsfield Unit (HU) values were processed using a minimum threshold of -200 and a maximum threshold of 300 prior to being normalized to yield values between 0 to 1. In order to focus on organs and suppress the background information, we cropped and reserved the regions of interest according to the body contour in the original CT images and use it as training data. Finally, to circumvent the computer memory limitation, data downsampling was performed using linear interpolation for CT images and using nearest interpolation for the labels. The image resolution in the LCTSC datasets was resampled to 2 mm x 2 mm x 2.5 mm. To further improve the segmentation performance of the esophagus and spinal cord, the volumes of these two organs were cropped separately for a further step we called "fine segmentation". The image resolution of these two organs was resampled to 1 mm x 1 mm x 2.5 mm. In the PCT dataset, the size of CT images was resampled to 144 x 144 x 144 pixels, which was found to generate better segmentation performance than the original resolution. Here in the PCT dataset, the training and resampling served to validate the general method proposed in this study. To achieve better computational efficiency, small organs in the PCT database were not subjected to the resampling process used for the LCTSC database.

## 2.A.2 Network architecture

The proposed network in this study is based on the 3D U-Net [29]. As shown in Fig. 2, the network consists of an encoder and a decoder. The encoder extracts image features while the decoder performs a voxel-level classification to achieve organ segmentation. The encoder contains 4 repeated residual blocks. Each block consists of 4 convolutional modules and each convolutional module is composed by a convolution layer with the kernel of 3x3x3, an instance normalization, and a leaky rectified linear unit. For each residual block, the stride of convolution layer in the convolutional modules is 1x1x1 except for the last convolutional module in which the stride is 2x2x2 to achieve the purpose of downsampling, and there is a spatial dropout layer between the early two convolutional modules to prevent the network from overfitting. The decoder contains 4 repeated segmentation blocks. Each block consists of 2 convolutional modules and 1 deconvolutional module. Four dashed arrows in the figure indicate four skipping connections that copy and reuse early feature-maps as the input to later layers having the same feature-map size by a concatenation operation to preserve high-resolution features. In the final 3 segmentation blocks, a 1×1×1 convolution layer is used to map the feature tensor to the probability tensor with the channels of the desired number of classes, n, before all results are merged by the upsampling operation to enhance the precision of segmentation results. Finally, a SoftMax activation is used to output a probability of each class for every voxel [40].

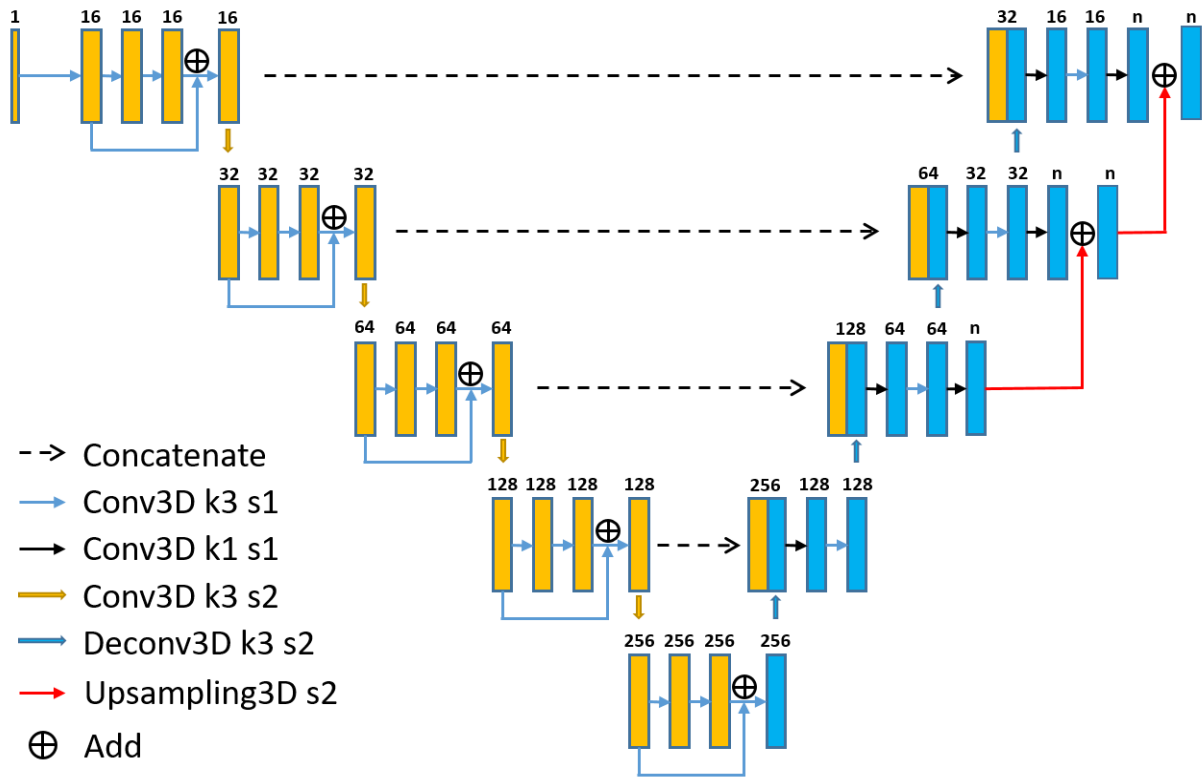

**FIG. 2.** The network architecture.

**2.A.3 Training**

For the LCTSC dataset, the 60 patients are divided into 5 groups with 12 patients per group. For the PCT dataset, the 43 patients are divided into 5 groups with 8 or 9 patients per group. Then the 5-fold cross-validation is performed. At the training stage, we first randomly extracted the patches from the resampled CT images to achieve data diversity and to prevent overfitting. The patch size is 96x96x96 in LCTSC and 128x128x128 in PCT. As noted earlier, the patch size for the fine segmentation of the esophagus and spinal cord in LCTSC is 32x32x32. Then, the network is trained by the patch and its corresponding labels. The loss function is defined as the weighted dice similarity coefficient as:

$$\text{Loss} = -\frac{1}{N*K} \sum_{i=1}^{N} \sum_{k=1}^{K} \frac{2*\sum_{v=1}^{V}(p_{i,k,v} * y_{i.k.v}) + \varepsilon}{\sum_{v=1}^{V} p_{i,k,v} + \sum_{v=1}^{V} y_{i,k,v} + \varepsilon},$$

where $p_{i,k,v}$ is the predicted probability of the voxel $v$ of the sample $i$ belonging to the class $k$, $y_{i,k,v}$ is the ground truth label (0 or 1), N is the number of samples, K is the number of classes, V is the number of voxels in one sample, and $\varepsilon$ is a smooth factor (set to be 1 in this study). The initial learning rate is 0.0005, and the Adam algorithm [41] is used to update the parameters of the network. The validation loss is calculated for every epoch, and the learning rate is halved when the validation loss no longer decreases after 30 consecutive epochs. The training process is terminated when the validation loss stops to decrease after 50 consecutive epochs.

**2.A.4 Testing**

In the testing stage, we first extract the patches from each CT images with a moving window with the size of 96x96x96 in LCTSC and 128x128x128 in PCT. The stride is 24 in LCTSC and 16 in PCT. In other words, we extract multiple patches from one patient and feed them into the network. The output of the network is a probability tensor for each patch. Then we merge all probability tensors from the same patient with a mean operator in the overlapping area to obtain the final probability tensor. Next, the class of each voxel is determined by the largest probability, which is the preliminary results of organ segmentation and the value of each voxel is the class number. Last, using the nearest neighbor interpolation, we resample the preliminary segmentation results to the size of original CT images to obtain the final organ segmentation results. It is worth noting that, for the fine segmentation of very small organs of esophagus and spinal cord in LCTSC, we first crop their own volume according to the multi-organ segmentation results, and then the volume is resampled to 1mm x 1mm x 2.5mm. Next, similar to the previous multi-organ segmentation, we extract the patches of 32x32x32 with a moving stride of 8x8x8 and separately feed them into their own networks. Finally, the fine segmentation of the esophagus and spinal cord were filled into the initial multi-organ segmentation results.

All experiments described above were performed on a Linux computer system. Keras with TensorFlow as the backend was used as the platform for designing and training the neural network [42]. The hardware includes: (1) GPU - Nvidia GeForce Titan X Graphics Card with 12GB memories, and (2) CPU - Intel Xeon Processor X5650 with 16GB

**2.B Organ dose calculations**

A GPU-accelerated Monte Carlo code, ARCHER, previously developed by members of this group was used in this study to calculate organ doses [20, 21, 43]. ARCHER is used in this study to simulate the transport of low-energy X-ray photons in heterogeneous media defined by the patient CT images where photoelectric effect, Compton scattering, and Rayleigh scattering can take place. CT scan protocols are predefined for ARCHER, including a combination of scan mode (helical or axial), beam collimation (5, 10, or 20 mm) and kVp (80, 100, 120 or 140). Fig. 3 illustrates the simulation environment involving a patient and a CT scanner.

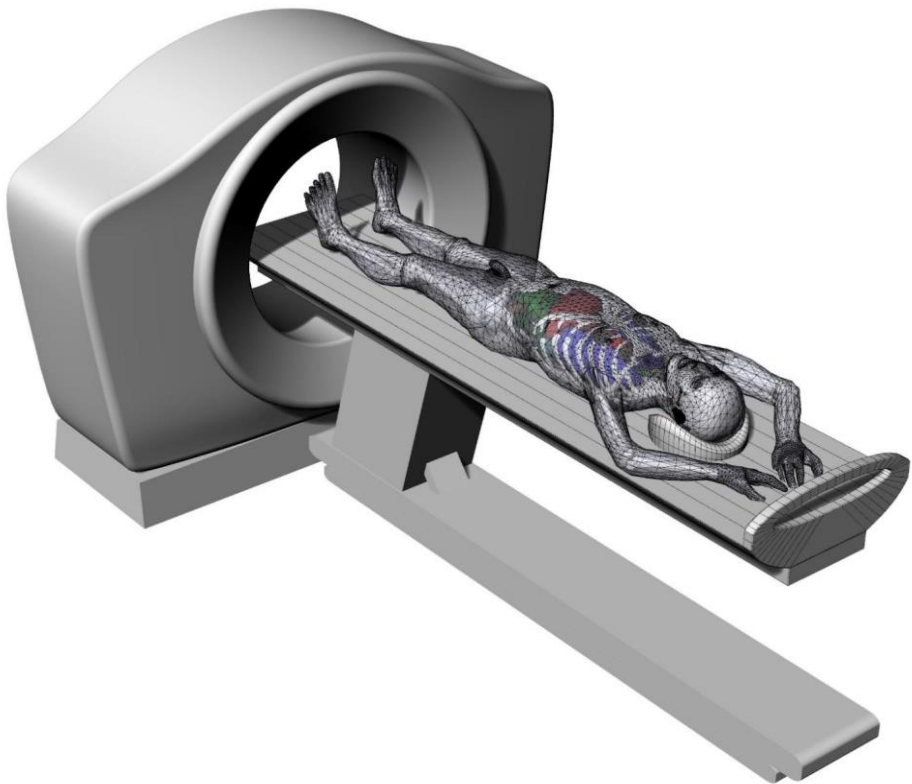

**FIG. 3.** CT dose simulation model of a patient undergoing a CT scan.

This study considered a scanner model representing a GE Lightspeed Pro 16 multi-detector CT that has been validated in our previous studies [44, 45], although a newer scanner model can be similarly created when needed in the future. The scanning protocol considered in this study includes 120 kVp, 20 mm beam collimation, axial body scan at a constant 100 mAs. CT scanner's continuous rotational motion is simulated using the step-and-shoot pattern, with each rotation approximated by 16 discrete positions [44]. Combining the newly segmented organ masks and voxel-wise dose maps calculated by ARCHER for a specific patient as is done in radiation treatment planning, we can derive the average absorbed dose for each organ of interest, as illustrated earlier in Fig 1. The computational speed is evaluated to make sure it is acceptable as part of the clinical workflow.

To show the potential clinical impact of the new method, patient-specific organ dose results are compared against organ doses derived from population-average phantoms used in the VirtualDose software [45, 46]. Fig.4 shows the RPI-Adult Male (73 kg in weight and 176 cm in height) and RPI-Adult Female (60 kg in weight and 163 cm in height) phantoms that were designed in accordance with anatomical parameters for the 50th percentile of the population [46]. When the weight and height of an adult patient are unspecified, the clinical organ dose assessment procedure at Massachusetts General Hospital usually picks these standard adult phantoms from the VirtualDose software to represent that patient.

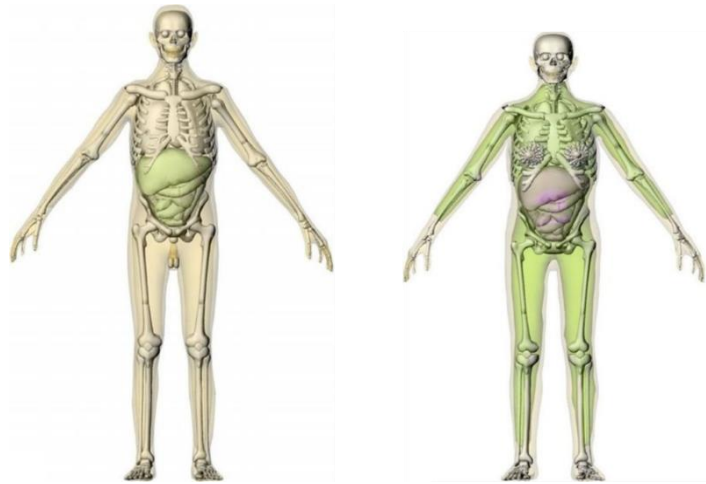

**FIG. 4.** RPI-Adult Male (left) and RPI-Adult Female (right) phantoms in the VirtualDose software that were designed in accordance with anatomical parameters for the 50th percentile of the population, thus bringing errors when compared with patient-specific organ doses [46]

**2.C Segmentation and organ dose evaluation criteria**

The Dice Similarity Coefficient (DSC) was used to evaluate the performance of organ segmentation [47]:

$$DSC = \frac{2|A \cap B|}{|A| + |B|}$$

where *A* is the volume of manually segmented organ (i.e., the ground truth) and *B* is the volume of the organ segmented by the network. The DSC ranges from 0 to 1 with the latter indicating a perfect performance. The Relative Dose Error (RDE) was used to evaluate the accuracy of dose calculation for each organ:

$$RDE = \frac{D - D_r}{D_r} * 100\%$$

where *D* is the organ dose calculated by ARCHER using patient-specific phantom consisting of either automatically segmented organs using the proposed the network (i.e., our method) or organs in the population-average phantom, and $D_r$ is the reference organ dose calculated by ARCHER using patient-specific phantom and manually segmented organ.

## 3. Results

### 3.A Organs segmentation

The performance of our network in organ segmentation is evaluated in terms of the DSC. As shown in Fig. 5, the segmentation results of all organs are summarized in these two box plots. For 60 patients from LCTSC, we achieved median DSCs of 0.97 (right lung), 0.96 (left lung), 0.93 (heart), 0.88 (spinal cord), and 0.78 (esophagus) as can be seen in Fig. 5 (a). For 43 patients from PCT, we achieved median DSCs of 0.96 (spleen), 0.96 (liver), 0.95 (left kidney), 0.89 (stomach), 0.87 (gall bladder), 0.79 (pancreas), 0.74 (esophagus), and 0.64 (duodenum) as can be seen in Fig. 5 (b). These values are very good and, as discussed later, meet the needs for CT organ dose reporting. Fig. 6 (a) and (b) show visual comparison of manual and automatic multi-organ segmentation results from both LCTSC and PCT, respectively, in axial, sagittal, coronal, and 3D views.

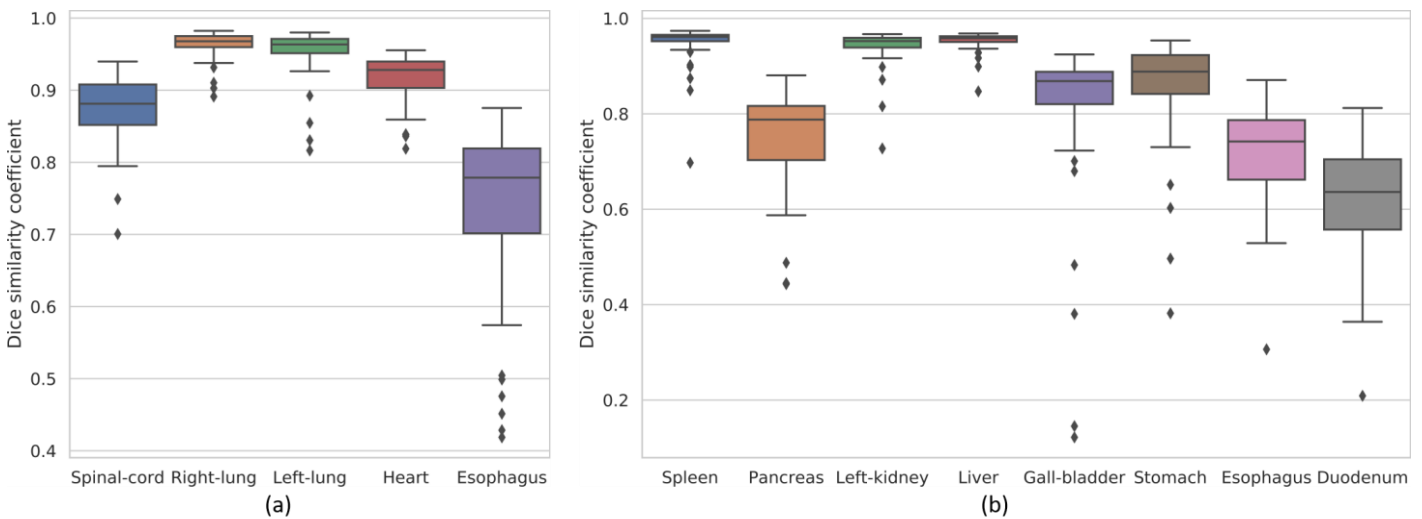

**FIG. 5.** Evaluation of organ segmentation performance in terms of DSC. (a) Data based on 60 patients from the LCTSC database. (b) Data based on 43 patients from the PCT database.

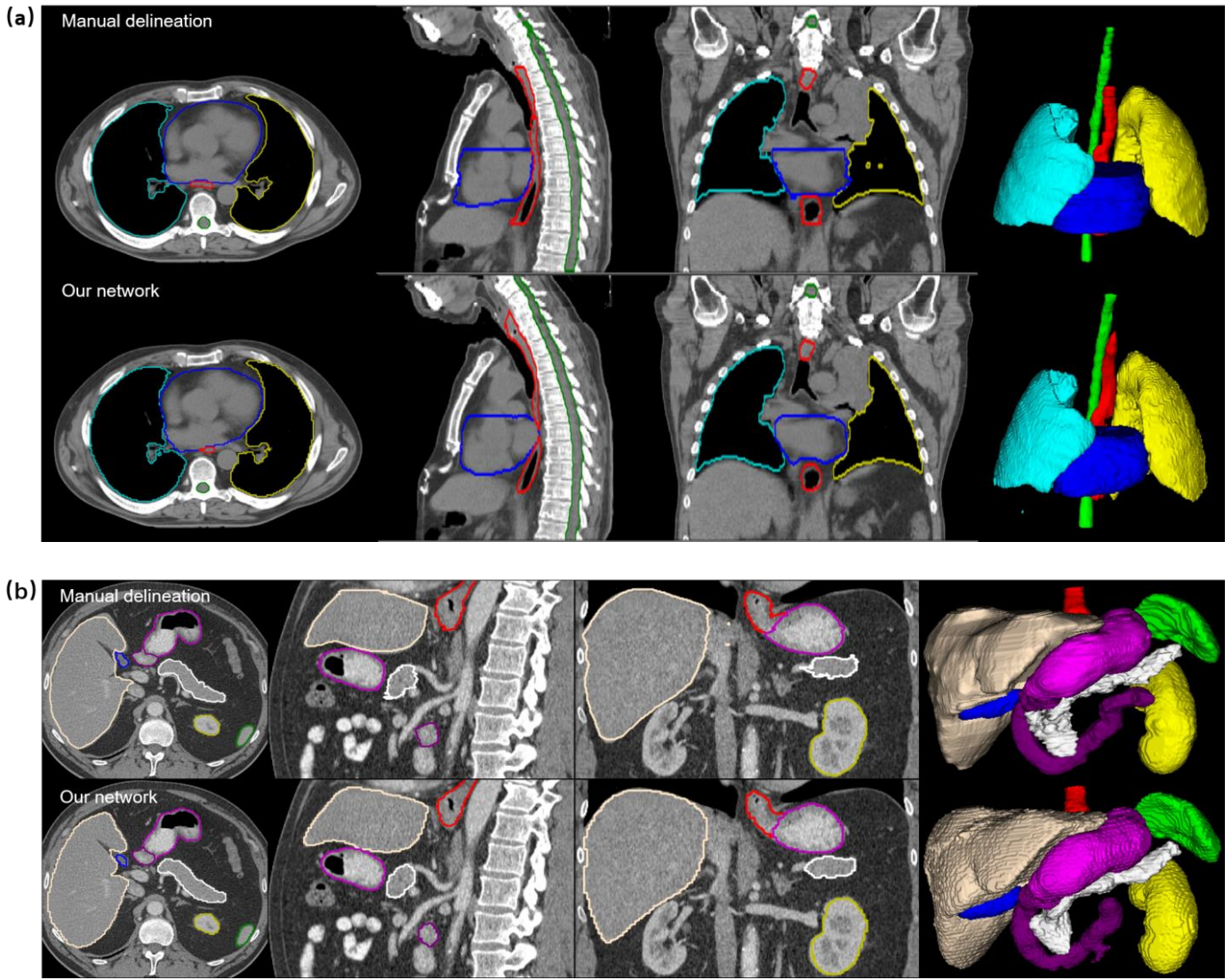

**FIG. 6.** Examples for visual comparison of organ segmentation between manual methods from LCTSC or PCT database (showed in the upper row in each panel) and our automatic method (showed in the upper row in each panel), in terms of axial, sagittal, coronal, and 3D views (from left to right). (a) LCTSC database showing left lung (yellow), right lung (cyan), heart (blue), spinal cord (green), and esophagus (red). (b) PCT database showing spleen (green), pancreas (white), left kidney (yellow), gallbladder (blue), esophagus (red), liver (bisque), stomach (magenta), and duodenum (purple).

The LCTSC database includes Reference DSC results provided by the organizer of the on-line challenge and DSC results submitted by 7 participating contestants who analyzed 12 patient CT datasets [37]. Table 1 summarizes the comparison of DCS results from this study with those from the LCTSC. The data show that our method has achieved comparable performance levels with those from the LCTSC database for the left lung, right lung, heart, and spinal cord. For the esophagus, which yielded worst performance among these five organs in the LCTSC database, our method scored better than all the 7 contestants, but ranks behind the Reference DSC result. These results suggest that our automatic organ segmentation method, which is used for purposes of CT organ dosimetry, has achieved the similar level of accuracy for the testing cases considered in this LCTSC database. Since the PCT database does not contain similar DSC data, Table 7 does not include organs in the abdominal region.

**TABLE 1**. Comparison of DSC results from this study with the Reference data and results from 7 contestants (DSC is given in "mean ± standard deviation").

| Methods | Left lung | Right lung | Heart | Esophagus | Spinal cord |
|---|---|---|---|---|---|
| LCTSC Reference | 0.96±0.02 | 0.96±0.02 | 0.93±0.02 | 0.82±0.04 | 0.86±0.04 |
| Method 1 | 0.97±0.02 | 0.97±0.02 | 0.93±0.02 | 0.72±0.10 | 0.88±0.04 |
| Method 2 | 0.98±0.01 | 0.97±0.02 | 0.92±0.02 | 0.64±0.20 | 0.89±0.04 |
| Method 3 | 0.98±0.02 | 0.97±0.02 | 0.91±0.02 | 0.71±0.12 | 0.87±0.11 |
| Method 4 | 0.97±0.01 | 0.97±0.02 | 0.90±0.03 | 0.64±0.11 | 0.88±0.05 |
| Method 5 | 0.96±0.03 | 0.95±0.05 | 0.92±0.02 | 0.61±0.11 | 0.85±0.04 |
| Method 6 | 0.96±0.01 | 0.96±0.02 | 0.90±0.02 | 0.58±0.11 | 0.87±0.02 |
| Method 7 | 0.95±0.03 | 0.96±0.02 | 0.85±0.04 | 0.55±0.20 | 0.83±0.08 |
| Our method | 0.96±0.01 | 0.96±0.02 | 0.93±0.02 | 0.75±0.10 | 0.88±0.04 |

**3.B Organ dose calculations**

The accuracy of organ dose calculations is evaluated in terms of RDE for the purposes of CT organ dosimetry where 10% is considered generally to be excellent. In the dataset from LCTSC, we calculated the organs doses of 60 patients in total, including the lung, heart, and esophagus. The left lung and right lung are treated as one organ, and the RDE of the spinal cord was not considered because it is not segmented in the population-average phantom. In the dataset from PCT, we calculated the organs doses of 43 patients in total, including the spleen, left kidney, gallbladder, liver, stomach, and pancreas. The duodenum is not segmented in the population-average phantom, and the esophagus in the specific patient is incomplete in the abdominal CT scanning, so the RDEs were not considered for the duodenum and esophagus. Results show that, comparing with the population-average-phantom based method (in parentheses next), our patient-specific method achieved the smaller RDE range for all organs: -4.3%~1.5% (vs -31.5%~33.9%) for the lung, -7.0%~2.3% (vs -15.2%~125.1%) for the heart, -18.8%~40.2% (vs -10.3%~124.1%) for the esophagus, -5.6%~1.6% (vs -20.3%~57.4%) for the spleen, -4.5%~4.6% (vs -19.5%~61.0%) for the pancreas, -2.3%~4.4% (vs -37.8%~75.8%) for the left kidney, -14.9%~5.4% (vs -39.9% ~14.6%) for the gallbladder, -0.9%~1.6% (vs -30.1%~72.5%) for the liver, and -23.0%~11.1% (vs -52.5%~-1.3%) for the stomach. These data are further compared in Fig. 7 (a) and (b) using box plots of RDEs for each organ in these two datasets. The ground-truth reference organ doses were calculated for the specific patient using ground-truth segmentation data from the original database. "Proposed method" represents the RDE between organ doses from our automatic segmentation and the reference organ doses, and "phantom-based method" represents the RDE between organ doses from the population-average phantom and reference organ doses.

As can been seen from data summarized earlier and the Fig. 7, the RDE values for our proposed patient-specific method has much smaller. In a CT scan, the height, weight, and organ topology of a patient can influence organ dose values. There is no doubt that it introduces some errors using population-average phantoms to replace a specific patient for organ dose calculation. In the case of dose to the heart, the current method of using population-average phantom in the VirtualDose software is found to suffer from the error range (-15.2% ~ 125.1%) due to the anatomical differences between the phantom and a real patient. As expected, the patient-specific method has a much smaller error range (-7.0% ~ 2.3%) for the heart, due to difference in organ segmentation between the CNN-based method and the ground truth. These results suggest that the patient-specific method can bring significant (in the case of dose to the heart, 125/7 times) improvement to the current CT organ dose assessment method that is based on population-average phantoms.

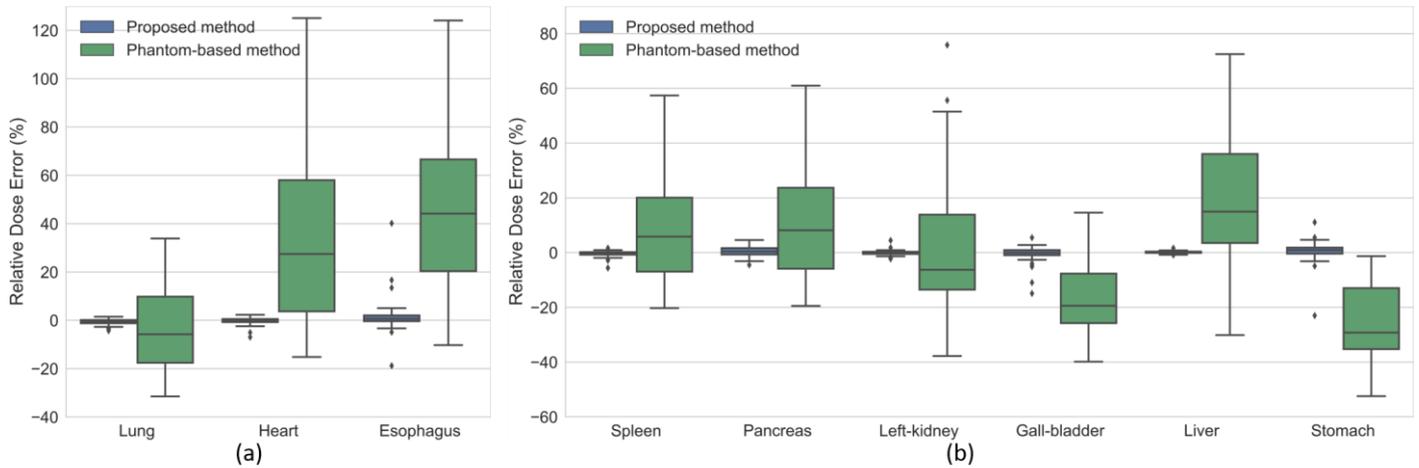

**FIG. 7.** The box plots of RDE showing that the proposed patient-specific method has much reduced errors when compared against the ground truth data. (a) For 60 patients from LCTSC and (b) For 43 patients from PCT.

### 3.C. Computational Efficiency

The computing time in our method, includes two processes performed in parallel as illustrated previously in Figure 1. The time for automatic organ segmentation for each patient is less than 5 seconds for all the 103 patient cases considered in the study. The time to calculate a total of $1 \times 10^8$ photons using ARCHER code running on an Nvidia Titan RTX GPU card with 24 GB memory is less than 4 seconds for all 103 patients with an organ dose statistical uncertainty of less than 0.5%. From our experiences, such computational accuracy and efficiency are expected to be acceptable as part of the routine clinical workflow.

### 3. DISCUSSION

In this study, we designed and trained a 3D CNN model to automatically segment thoracic and abdominal organs in patient-specific CT images using two publically available databases. In particular, we developed a fine segmentation procedure to treat narrow and long organs such as the esophagus and spinal cord in the LCTSC database and improved their segmentation performance. In the duodenum, like the other on-line test participants, the segmentation performance of our network was relatively poor, because the organ and its surrounding tissues have similar pixel values in CT image making the boundary difficult to detect by the CNN model. Nevertheless, results from this study have clearly demonstrated the accuracy and efficiency of the CNN model in performing the automatic multi-organ segmentation task for the purposes of assessing patient organ doses. Implementation of the proposed method can lead to significant improvement in the accuracy of organ dose calculation based on the population-average phantoms.

It is important to note that the significance of this study is not about the ability to perform CNN-based automatic multi-organ segmentation. As evidenced by the 2017 AAPM Thoracic Auto-Segmentation Challenge, start-of-the-art automatic segmentation methods, DL-based or atlas-based, can already achieve impressive performances [37]. Therefore, the objective of this study was never to invent a new and better organs segmentation method. Instead, the significance of this study is that, for the first time, we have demonstrated that it is feasible to combine DL-based automatic multi-organ segmentation tool with the GPU-based rapid Monte Carlo dose calculation code in a streamlined process that takes less than 5 seconds for each patient. With this newly demonstrated capability of "patient-specific" organ dose assessment, future CT scanners can take advantage of patient- and scan-specific features in a new paradigm – the "prospective" design of tube voltage and current modulation, beam collimation and filtering, and gantry angle – leading to the ultimate goal of achieving low-dose and optimized CT imaging.

There are several limitations in the current study. The variable and somewhat less accurate performance of our approach for segmenting narrow and long structures with poor soft-tissue contrast such as the esophagus and duodenum may be related to the relatively small size of training data in the databases causing irregularities in CT attenuation and position of these structures. Another limitation was that we did not assess the effect of major abnormalities on the organ segmentation – an issue already recognized by organizers of the 2017 Lung CT Segmentation Challenge [37]. Likewise, diffuse abnormalities and paucity of intra-abdominal fat can have a negative effect on the ability of our segmentation algorithm. Further studies should consider larger patient data size and cover children and additional radiosensitive organs in the head and neck regions. One set of unique data already available from MGH is the annotated cadaver CT images that can be ideally for testing of DL-based image analysis and dosimetry algorithms [48-50].

Finally, it is worth noting that, with the patient-specific organ dose information, one can derive the so-called "effective dose" – a quantity that the American Association of Physicists in Medicine (AAPM) believes to bear significant uncertainty and therefore should be used only for prospective radiologic protection purposes and to help patients understand medical radiation dose in perspective [51].

## 5. CONCLUSION

In this study, an automatic multi-organ segmentation method was developed using a CNN model that was trained with two publically available CT databases involving a total of 103 patients. The method takes less than 5 seconds to perform automatic multi-organ segmentation for a patient and, for purposes of CT organ dosimetry, has achieved the similar level of segmentation accuracy for the testing cases considered this study. The organ dose calculation method takes less than 4 seconds for a total of $1\times10^8$ photons using the GPU-based rapid Monte Carlo code, ARCHER, to achieve the organ dose statistical uncertainty of better than 0.5%. These results demonstrate, for the first time, the excellent accuracy and efficiency of a streamlined patient-specific organ dosimetry computational tool. Implementation of such methods as part of the clinical workflow can yield considerable improvement over the current CT organ dose methods that are based on population-average phantoms, thus opening the door to prospective patient-specific optimization features in the future.

**CONFLICT OF INTEREST DISCLOSURE**

X.G. Xu is co-founder of Virtual Phantoms, Inc (Albany, New York) that commercializes software technologies—VirtualDose™ for medical dose reporting and ARCHER™ for real-time Monte Carlo dose computing funded by grants from NIH/NIBIB (R42EB010404 and R42EB019265-01A1).